\documentclass[twocolumn, superscriptaddress, prl, amsmath, amssymb, aps, 10pt, floatfix]{revtex4-2}

\usepackage{float}
\usepackage{graphicx}
\usepackage{dcolumn}
\usepackage{bm}
\usepackage{xcolor}
\usepackage{textgreek}
\usepackage{textcomp}
\usepackage[separate-uncertainty=true, multi-part-units=single, list-units=repeat]{siunitx} 
\usepackage[version=4]{mhchem} 
\usepackage[thinc]{esdiff} 
\usepackage{comment}

\begin{document}

\preprint{APS/123-QED}

\title{ No time for surface charge: how bulk conductivity hides charge patterns from KPFM in contact-electrified surfaces}

\author{Felix Pertl}
  \email{felix.pertl@ist.ac.at}
\affiliation{Institute of Science and Technology Austria, Am Campus 1, 3400 Klosterneuburg, Austria}
\author{Isaac C.D. Lenton}%
\affiliation{Institute of Science and Technology Austria, Am Campus 1, 3400 Klosterneuburg, Austria}
\author{Tobias Cramer}
\affiliation{Department of Physics and Astronomy
University of Bologna,
Viale Berti Pichat 6/2, 40127 Bologna, Italy}
\author{Scott Waitukaitis}
\affiliation{Institute of Science and Technology Austria, Am Campus 1, 3400 Klosterneuburg, Austria}

\date{\today}

\begin{abstract}

Kelvin probe force microscopy (KPFM) is widely used in stationary and dynamic studies of contact electrification. An obvious question that connects these two has been overlooked: when are charge dynamics too fast for stationary studies to be meaningful? Using a rapid transfer system to quickly perform KPFM after contact, we find the dynamics are too fast in all but the best insulators. Our data further suggests that dynamics are caused by bulk as opposed to surface conductivity, and that charge-transfer heterogeneity is less prevalent than previously suggested.

\end{abstract}

\maketitle
Contact electrification (CE), \textit{i.e.}~the transfer of electrical charge when materials touch, occurs in settings ranging from coffee grinding \cite{harper2024moisture}, to pollen transport \cite{montgomery2021bumblebee} and perhaps even rocky planet formation \cite{steinpilz2020electrical}, yet is poorly understood \cite{Lacks.2019, lacks2012unpredictability}. Among the most useful tools for studying CE are so-called `scanning Kelvin' techniques, which enable imaging of voltages caused by transferred charge. At the macroscale ($\sim$100 \textmu m to 10 cm), the main method is `scanning Kelvin probe microscopy' (SKPM), while at the nanoscale ($\sim$10 nm to 100 \textmu m) the most used is the related (but distinct) `Kelvin probe force microscopy' \cite{zisman1932new, craig1970stress, nonnenmacher1991kelvin, melitz2011kelvin, pertl2022quantifying, lenton2024beyond}. Both methods involve slowly scanning a metal tip over the surface while recording a voltage. Depending on the size and resolution of a scan, this can take from minutes to even hours. These methods have led to two primary veins of research with implications for CE. In the first, they have focused on \textit{stationary} patterns of charge left after CE, which are occasionally observed to be heterogeneous \cite{Baytekin.2011, Shinbrot.2008, Burgo.2012, Moreira.2020ikk, Knorr.2011, terris1989contact, Barnes.2016, Sobolev.2022, hull1949method, bertein1973charges, ji2021stability, Gonzalez.2017}. In the second, the focus has been on the \textit{dynamics} of charge deposited by CE, \textit{i.e.}~how it evolves over space and time, where proposals have included surface diffusion, surface drift, bulk drift, and combinations thereof \cite{Baytekin.2011,Bai.2021, navarro2023surface, Knorr.2010}. Considering these two parallel veins of research, an obvious question arises: under what circumstances are the \textit{dynamics} too fast for measured \textit{stationary} patterns to be meaningful?

In this work, we address this overlooked question that connects the stationary and dynamic studies of CE with Kelvin probe techniques. We perform CE with macroscopic (1$\times$1 cm$^2$) samples and then quickly (order one minute) image the surface with KPFM. We find that CE deposited charge is (1) largely uniform on the surface over KPFM length scales and (2) leaves in a manner that appears as time-decay in the KPFM voltage. Based on these observations, we model the charge dynamics as the discharge of a simple capacitor, where the time-constant is set by a material's permittivity and electrical conductivity. To further support this model, we extend our experiments to a wide variety of `good insulators' spanning nearly four orders of magnitude in nominal conductivity, showing that the better the insulator, the longer the decay. Our results call into question the validity of stationary studies of CE with KPFM on all but the best electrical insulators.

\begin{figure}[!t]
\centering
\includegraphics{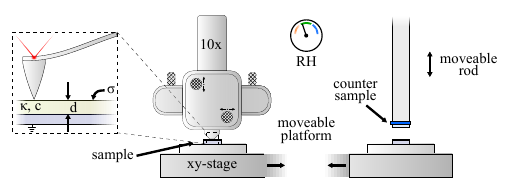}
\caption{ \label{fig:physical_setup} Experimental setup. We use a commercial AFM equipped with a conductive cantilever in non-contact mode to simultaneously capture topography and surface potential (KPFM) measurements. The `main' (planar and insulating) sample is fixed onto a grounded platform, which is moveable by two sets of stages: (1) a piezo stage for fine positioning during KPFM measurement, and (2) a macro-stage for long-distance positioning. The latter allows us to quickly ($\sim$10 s) move the sample below a second setup, where it is contacted by a `counter sample' on a vertical linear actuator. In a typical experimental protocol, we begin with a KPFM measurement of the main sample in its discharged state. The sample is then moved beneath the (typically PDMS) counter-sample, where it is contacted with a set pressure. The main sample is then returned to its original position for a post-contact KPFM measurement in the same region. This process is completely automated, making it repeatable over multiple cycles.}
\end{figure}

The experimental setup is illustrated in Fig.~\ref{fig:physical_setup}. A key feature is the incorporation of a macro-stage to move a sample between the AFM (where KPFM is performed) and second setup that uses a linear actuator to execute charge-exchanging contacts with a counter sample. While in a typical system the sample transfer, AFM approach, re-initiation and re-calibration of the KPFM parameters can easily take as long as tens of minutes, in our system this happens in as little as $\sim$30 s. We use a variety of materials for the `main' sample, but always use polydimethylsiloxane (PDMS) for the counter sample as its softness and smoothness enable it to make `conformal' contact over an entire main sample ($\sim$$1$$\times$$1$ cm$^2$). Hence, we can perform KPFM scans on multiple regions of an uncharged sample, move it underneath the counter sample to perform charge-exchanging contacts, and then return it to the AFM for KPFM scans after CE has occurred, nominally at the same positions. As all aspects of the system are completely automated, we can repeat this process many times. Our AFM (NX20, Park Systems) is equipped with a conductive cantilever (NSC14/Cr-Au, Mikromasch), and enables us to obtain surface topography and potential simultaneously in non-contact mode. The whole setup is housed in a ISO class 5 cleanroom with rigid temperature (21.1$\pm$0.2)$^{\circ}$C and humidity (43$\pm$4)\% regulation. Our thin insulator samples are prepared on gold-coated Si-wafers, which act as the constant-potential back electrode required for the KPFM measurement. Sample thicknesses range from approximately one to a few hundred microns, depending on the material and as constrained by the compensation range ($\pm$10 V) of the AFM.
For details about the sample preparation and properties, see Supplemental Material \cite{SupplMat}. All measurements are done in amplitude modulation mode (AM-KPFM).

\begin{figure}[!t]
\centering
\includegraphics{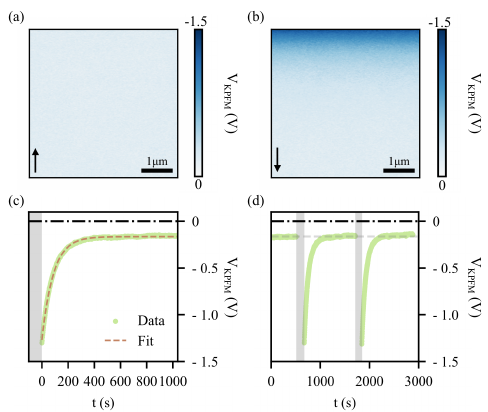}
\caption{ \label{fig:pdms_decay} Unexpected spatial gradients reveal time dependence. We conduct KPFM scans (5$\times$5 {\textmu}m$^2$) on a $\sim$100 {\textmu}m thin spin-coated PDMS layer atop a gold-coated Si wafer. (a) Before contact, the KPFM potential is spatially uniform, with a mean value of $-0.17 \pm 0.03$ V. (b) After contact with another PDMS counter sample, we observe spatial gradient in the potential, which decreases in magnitude along the slow scan direction (indicated by the arrow). (c) By averaging the potential across each line and plotting it over time, we see that this decay is not due to `real' spatial variability, but rather some time-dependent process. (d) Repeating the contact and measurement procedure reveals a reproducible, time-dependent potential decay with the same characteristics; notably the spatial gradient is \textit{always} in the slow scan direction.}
\end{figure}

Our efforts in this project started as an attempt to reproduce data that exists elsewhere in the literature. Specifically, we intended to study contact electrification between two nominally identical PDMS samples, which in previous studies \cite{Baytekin.2011} resulted in KPFM maps with heterogeneous features of positive/negative voltage alternating over a lateral scale of a few hundred nanometers. In our protocol, we begin by fully discharging both the main and counter sample (see Supplemental Material for more information) and then performing KPFM on the main sample \cite{SupplMat} (Fig.~\ref{fig:pdms_decay}(a)). In this state, we observe a spatially uniform potential, similar to what was reported before \cite{Baytekin.2011}. We then use the macro-stage to move the main sample below the counter sample and perform one charge-exchanging contact. After returning the original position, we again measure a KPFM map. Without exception, the KPFM measurement after contact does not exhibit alternating regions of plus/minus polarity, but instead a spatial gradient of a single polarity (Fig.~\ref{fig:pdms_decay}(b)). Tellingly, this spatial gradient is always perfectly aligned with the slow-scan direction of the AFM (indicated by the black arrow in the figure). This suggests it is rather a signature of time-dependence as opposed to a space-dependence. If we unfold the KPFM data and plot it against measurement time instead of position, we observe a smooth, ostensibly exponential curve, which returns to the pre-contact voltage about 400 seconds after contact, shown in Fig.~\ref{fig:pdms_decay}(c). By repeating the KPFM/contact/KPFM procedure, we observe this decay over and over again (Fig.~\ref{fig:pdms_decay}(d)). 

What causes this time dependence? Our thinking is as follows. First, we imagine that in the instant just after contact, the CE-deposited charge (and hence KPFM potential) is not heterogeneous, but instead quite uniform---at least over a lateral scale large compared to our sample thickness of the KPFM scan region. With this assumption, we approximate the insulator surface and conductive back electrode as forming a parallel plate capacitor. The uniform charge density, $\sigma$, on top of the insulator surface creates an electric field, $\vec{E}$, inside the bulk. If the insulator were perfect---\textit{i.e.}~with zero electrical conductivity---nothing interesting would happen. The charge on the top surface would remain constant and the spatially uniform KPFM signal would simply be equivalent to the voltage across the capacitor. Yet perfect insulators do not exist, hence over some timescale charge flows. Assuming instead that our sample is just a `good insulator', and furthermore one that exhibits Ohmic electrical conductivity, we expect the electric field to drive a current density in the bulk given by
\begin{equation}
    J = c E = c \frac{\sigma}{\kappa \epsilon_0} = \dot{\sigma},
    \label{eq:Ohms law}
\end{equation}
where $c$, $\kappa$ and $\dot{\sigma}$ are the electrical conductivity, relative permittivity and charge rate over time, respectively. This relationship results in an exponential decay in the surface charge density, which in turn predicts a decay in the voltage, given by
\begin{equation}
    V_{KPFM}(t) = V_{bg} + V_0 e^{-t/\tau},
    \label{eq:voltage decay}
\end{equation}
where $V_{bg}$ is the `background potential' that exists in the absence of charge, $V_0$ is the change in the initial potential due to the added CE charge, and $\tau = \kappa \epsilon_0/c$ is the characteristic time constant. This model implies that the decay is governed by screening from mobile bulk charges of the material, not by movement of the deposited charge on the surface.

\begin{figure}[!t]
\centering
\includegraphics{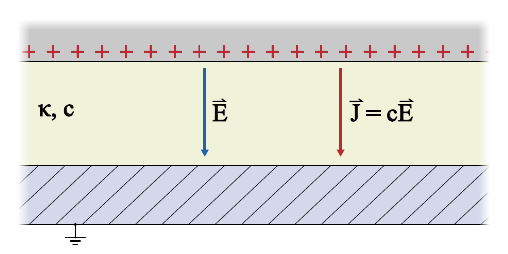}
\caption{ \label{fig:capacitor_model} A simple model for bulk charge decay. We imagine that the insulator sample and the grounded substrate form a parallel-plate capacitor, where the charge added to the surface during contact electrification creates a potential difference across the bulk. Owing to the finite conductivity of the bulk, $c$, this drives a current density, $\Vec{J}$, between the surface and ground.
}
\end{figure}

\begin{figure*}[!htbp]
\centering
\includegraphics{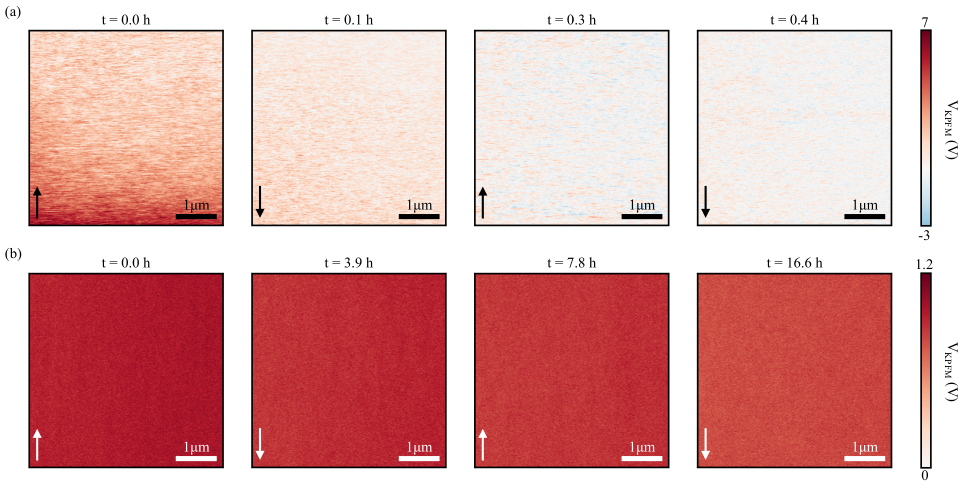}
\caption{ \label{fig:long_decay} Predicting the behavior of other materials. We test our model by using other materials with different values of bulk conductivity, with the prediction that materials with lower conductivities should decay more slowly. (a) KPFM scans of SU-8 after contact by PDMS, which exhibits a decay over a slightly longer timescale than PDMS. (b) KPFM scans fo SiO$_2$ after contact by PDMS, where decay is only apparent after tens of hours.}
\end{figure*}

At first glance on a linear scale, Eq. \ref{eq:voltage decay} seems to fit the data in Figs.~\ref{fig:pdms_decay}(c,d) well, revealing a time constant of $\sim$90 s. If we look up the nominal values for the dielectric constant ($\sim$2.72) and conductivity ($\sim$$3.45\times10^{-13}$ S/m) of PDMS, our model predicts a decay timescale of $\sim$70 s; not bad. However, it must be pointed out that such literature values are indeed only nominal, due to the fact that `good insulators', like PDMS, are generally non-Ohmic. Indeed, we show in the Supplemental Material that more careful examination of the decay on a log-lin scale exhibits non-exponential features, which fit better to more complex models \cite{molinie1995surface,molinie1996potential,molinie2023dielectric}, \textit{e.g.} the Cole-Cole response function \cite{SupplMat}.

Explaining such non-Ohmic conductivity is itself a challenging topic of ongoing theoretical and experimental investigation \cite{molinie1995surface, molinie2023dielectric}, and lies far outside the focus of this work. Instead, our focus remains what we stated previously: can the dynamics of CE charge on a surface be too fast for stationary KPFM measurements to be meaningful? Continuing with this aim, we perform additional experiments with different materials. Fig.~\ref{fig:long_decay}(a) shows KPFM scans taken after a PDMS counter sample has contacted an SU-8 main sample, which has a nominal conductivity that is somewhat lower (nominally 10-100$\times$). Indeed, we again observe a decay, and one that occurs over a longer timescale. To slow down the decay even further, we use a thermally grown SiO$_2$ oxide layer, which has a nominal conductivity that is substantially lower (somewhere in the range of 100-1000$\times$). Now it is not easy to see any time dependence in individual scans. However, spacing out measurements over the same region by hours and even days, we tease out very slow changes to the potential, as shown in Fig.~\ref{fig:long_decay}(b). It is worth mentioning that, for both of these materials, we again observe no signatures of charge heterogeneity. In the SU-8 data, the decay is still fast enough to see during the timescale of a single KPFM scan, but the voltage is still of a single polarity, consistent with our assumption that the potential is spatially uniform over a relatively large scale. In the SiO$_2$ data, we see this outright; the decay timescale is so much longer than a single scan that we can trust our eyes that the sign of charge on the surface is spatially homogeneous. 

In the Supplemental Material, we present analogous data for several other materials of varying nominal conductivities \cite{SupplMat}. Every material exhibits time-dependent decay, which is generally longer for better insulators and shorter as nominal conductivities increase.  Furthermore, all KPFM potentials we have observed are of a single polarity, and therefore CE in every case appears consistent with charge exchange of a single sign over space. We emphasize again that pretending these materials are Ohmic is a crude assumption; in reality, they are non-Ohmic, hence their decays are non-exponential (see Supplemental Material \cite{SupplMat}), and hence none of them can be ascribed a single value of conductivity. If, for instance, one looks up literature values for the `conductivity of SiO$_2$', one finds measurements ranging from 10$^{-13}$ to 10$^{-16}$ S/m \cite{palumbo2020review, shackelford2000crc}. Nonetheless, our data shows that indeed `better insulators' tend to decay more slowly, and this has implications for what types of materials KPFM-based CE experiments can be meaningful. 

Can we exclude other mechanisms that might cause the decays we observe? Several studies in the literature \cite{Bai.2021, navarro2023surface, im2023enhanced, mirkowska2014atomic} report lateral spreading on CE-charged insulator surfaces, but significant geometric differences separate our experiments from those. We use a large ($\sim$$1\times1$ cm$^2$ square), conformal counter sample to charge the entire surface of our main sample, and our data is consistent with charged regions whose lateral extent is much larger than the sample thickness. Experiments where lateral spreading is observed almost exclusively correspond to charged regions whose lateral extent is comparable to or even much smaller than the sample thickness. Often, such small spots are created by using the AFM to `scratch' a region (\textit{e.g.}~a $1\times1$ {\textmu}m$^2$ square) of charge onto the surface, and then lifting the tip off to subsequently image it with KPFM. While lateral spreading has been attributed to charge diffusion \cite{Bai.2021}, recent analysis has shown its more likely due to the electric field created by the spot driving surface and bulk conduction \cite{navarro2023surface}. Moreover, extracting the relative contributions of these by fitting to a coupled model, the bulk seems to have the (significantly) larger effect \cite{molinie2023dielectric}. 

\begin{figure}[!t]
\centering
\includegraphics{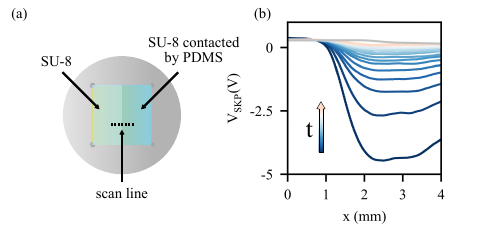} 
\caption{ \label{fig:skpm_experiments} Macroscopic observation of decay. (a) To be sure that the observed decay is primarily due to bulk, as opposed to surface, conductivity, we carry out macroscopic surface potential measurements with SKPM. We attach an SU-8 sample onto a gold-coated Si wafer. (b) Before contact, line scans at the same position yield a stable SKPM potential near zero $(0.25 \pm 0.05)$ V, shown as in gray. Contacting half of the SU-8 surface creates a change to about -5 V. Repeated scans over the same region over several hours reveal a time-dependent decay of the SKPM potential back to its initial value. As is visually evident, this decay occurs without lateral spreading, indicating bulk conductivity is the dominant mechanism.}
\end{figure}

To allay suspicion that lateral spreading accounts for our observations, we probe CE at the sample (centimeter) length-scale with SKPM. We prepare a fully discharged, SU-8 sample and begin by performing 1D scans across the center, as indicated by the dashed line in Fig.~\ref{fig:skpm_experiments}(a). These reveal an essentially flat and nearly zero potential (gray line in Fig.~\ref{fig:skpm_experiments}(b)).
Next, we contact half of the SU-8 sample with a PDMS counter sample, and then remeasure the potential repeatedly over the same scan line. There are three noteworthy observations in this data. First, as we hypothesized based on the KPFM data, the polarity of the signal is homogeneous---not heterogeneous---indicating spatially uniform charge transfer during CE. Second, the signal decays with time, as indicated by the arrow. Moreover, this decay occurs without any significant lateral spreading \cite{haenen1976experimental}, strongly suggesting that the observed decay is governed by bulk conductivity rather than surface mechanisms.

In summary, our data leads to several meaningful conclusions. First, a practical but nonetheless very important point that has been overlooked: KPFM is useful for addressing stationary patterns of CE-transferred charge only in the very best insulators. This is most strikingly illustrated by comparing the cases of PDMS and SiO$_2$. In PDMS, CE-transferred charge is screened so quickly that its decay is visible during a single KPFM scan. If we did not have our special system that allows us to quickly do KPFM after contact, it would be gone \textit{before} the scan. Hence, stationary studies with insulators of comparable resistivity to PDMS may very well be showing the `leftovers' of what remains after all CE-deposited charge has been effectively screened. In SiO$_2$, on the other hand, CE-deposited charge is stable over a timescale much longer than a single KPFM scan or the time required to transfer a sample. Hence, stationary studies with SiO$_2$ and other `top tier' insulators are (much) more likely to be meaningful. Second, all signs in our work point towards charge decay being dominated by bulk conduction. A more careful analysis than ours would consider non-Ohmic conduction, but that field of study lies outside of our interest. Third, while it has been suggested that charge-transfer in CE is inherently heterogeneous, our data shows that this is not at all certain. All of our data is consistent with largely homogeneous charge transfer, suggesting that when heterogeneity does occur, it is most likely due to a secondary effect (\textit{e.g.}~`sparks' after primary CE has occurred \cite{Sobolev.2022}).

This project has received funding from the European Research Council (ERC) under the European Union’s Horizon 2020 research and innovation programme (Grant agreement No.~949120). This research was supported by the Scientific Service Units of The Institute of Science and Technology Austria (ISTA) through resources provided by the Miba Machine Shop and the Nanofabrication Facility.

\bibliographystyle{apsrev4-2}
\bibliography{manuscript}

\end{document}


\setcounter{figure}{0}
\renewcommand{\figurename}{SUPPL.~FIG.}
\renewcommand\thefigure{\arabic{figure}}
\setcounter{table}{0}
\renewcommand{\tablename}{SUPPL.~TABLE}
\renewcommand\thetable{\arabic{table}}
\renewcommand{\theequation}{S\arabic{equation}}
\setcounter{equation}{0}  

\preprint{APS/123-QED}

\title{No time for surface charge: how bulk conductivity hides charge patterns from KPFM in contact-electrified surfaces}

\author{Felix Pertl}
  \email{felix.pertl@ist.ac.at}
\affiliation{Institute of Science and Technology Austria, Am Campus 1, 3400 Klosterneuburg, Austria}
\author{Isaac C.D. Lenton}%
\affiliation{Institute of Science and Technology Austria, Am Campus 1, 3400 Klosterneuburg, Austria}
\author{Tobias Cramer}
\affiliation{Department of Physics and Astronomy
University of Bologna,
Viale Berti Pichat 6/2, 40127 Bologna, Italy}
\author{Scott Waitukaitis}
\affiliation{Institute of Science and Technology Austria, Am Campus 1, 3400 Klosterneuburg, Austria}

\date{\today}


\maketitle

\section{Sample preparation and characterization}
\textbf{PDMS samples:} As the conducting substrate for a main sample,  we used used a gold coated Si-wafer cut it into a square of 1 cm side length. We prepared PDMS by mixing Sylgard™ 184 elastomer base and curing agent in a 10:1 ratio. After thoroughly blending with a centrifugal, bubble-free mixer (Hauschild SpeedMixer DAC 150.1, two minutes at 2000 rpm), we spin coated the uncured liquid at 1000 rpm for 1 min onto the gold surface. Baking this in the oven at 80$^{\circ}$C for 24 hours resulted in our `main' PDMS sample with a typical thickness of $\sim$100 \textmu{m}. This method was used for all data discussed in the main text.

In Suppl.~Fig.~1, we present additional data (`PDMS 65$^{\circ}$C') prepared identically, but cured at 65$^{\circ}$C instead.

\textbf{SU-8 samples:}  We spin-coated SU-8 (GM1075, Gersteltec) onto a gold-coated silicon wafer at 1000 rpm for 1 minute. This was subsequently baked on a hot plate at 150$^{\circ}$C for 10 minutes.  Typical sample thickness were $\sim$100 \textmu{m}.

\textbf{Mica samples:} We used a Mica disk (Agar Scientific) and cleaved it with a razor blade until we reached a thin flat sheet. Typical sample thickness were $\sim$50 \textmu{m}.

\textbf{Polyacrylonitrile samples:} Polyacrylonitrile (PAN) was deposited onto a gold wafer with an RF sputtering system (DST3-T, VacTechniche) under argon flow for several hours. Typical sample thicknesses were tens of nanometers.

\textbf{SiO$_2$ samples:} The SiO$_2$ wafer was obtained from a commercial distributor with a thickness of 3 \textmu{m}.

\textbf{PDMS counter samples:} When making main PDMS samples, remaining uncured elastomer was poured into a plastic dish, degassed until no bubbles remained and baked for 24 hours at 80$^{\circ}$C. From this slab, we cut out counter-samples of PDMS with 1 cm side length, using a custom-fabricated stencil to precisely guide the blade of a clean, fine razor. We adhered each PDMS counter sample to a 3D-printed holder using additional PDMS and further curing at 80$^{\circ}$C for 24 hours. Typical counter samples had a thickness of a few millimeters. We used the `air-facing' side of these samples for contact. 

We used a second method to make PDMS counter samples following Ref.~\cite{Baytekin.2011}. Instead of curing the PDMS in a plastic dish, we cast it onto a silanized Si-wafer (1H, 1H, 2H, 2H-perfluorooctyltrichlorosilane) and baked it at 65$^{\circ}$C for 24 hours. We gently peeled off the PDMS from the wafer and cut it into squares. These were rinsed with dichloromethane (DCM) and isopropyl alcohol (IPA) and dried with nitrogen to remove any residue of the silane. We adhered each sample to a 3D-printed holder using additional PDMS, and then let this cure at room temperature for at least 48 hours. We used the `wafer-facing' side of these samples for contact.

\textbf{Sample thickness measurements:} The thicknesses of PAN and SiO$_2$ samples were measured with an ellipsometer (Multi-Wavelength, Film-Sense). All other samples were measured using the AFM by bringing the cantilever into contact with the sample on the sample stage, then subtracting the z-scanner position, as detailed in Suppl.~Table \ref{tab:material_properties}. 

\textbf{Sample discharge:} We used a custom-built discharge chamber to remove residual charge from the sample before contact. The chamber housed a photoionizer (Hamamatsu L12645, 10 keV) positioned in front of a fan. Samples were placed downstream in the airflow generated by the fan. As the photoionizer ionized the air, the resulting charged species were carried toward the sample. The sample discharged as its electric field attracted oppositely charged ions.

\begin{table*}[ht!]
\centering
\caption{\parbox[t]{\textwidth}{\raggedright Properties of materials used:  thickness $d$, permittivity $\kappa$ and bulk conductivity $c$; $^{\dagger}$provided by distributor. \newline
$^{\ddagger}$ Average of minimum and maximum value. \newline \vspace{1mm}}}
\label{tab:material_properties}
\begin{tabular}{lccccccc}

\hline\hline
Material & d ($\mu$m) & $\kappa$ & $\kappa_\mathrm{max}$ & $\kappa_\mathrm{min}$ & $c$ (S/m) & $c_\mathrm{max}$ (S/m) & $c_\mathrm{min}$ (S/m) \\
\hline
SU-8         & $\sim$200 & 3.2$^{\dagger}$   & 4.1\cite{su8kayakuupperkappa}   & 2.85 \cite{ghannam2009dielectric}  
& $10^{-14}$$^{\dagger}$  & $5 \times 10^{-16}$\cite{tijero20098} & $1.28 \times 10^{-13}$\cite{su8kayaku} \\

PDMS 65$^{\circ}$C & $\sim$100 & 2.72$^{\dagger}$  & 3.0\cite{cresson20141}   & 2.55\cite{cresson20141}   
& $3.45 \times 10^{-13}$$^{\dagger}$ & $2.5 \times 10^{-14}$\cite{haynes2016crc} & $10^{-12}$\cite{xu2010micropatternable} \\

PDMS 80$^{\circ}$C & $\sim$100 & 2.72$^{\dagger}$  & 3.0\cite{cresson20141}    & 2.55\cite{cresson20141}    
& $3.45 \times 10^{-13}$$^{\dagger}$ & $2.5 \times 10^{-14}$\cite{haynes2016crc} & $10^{-12}$\cite{xu2010micropatternable} \\

Mica         & $\sim$57  & 8.1\cite{weeks1922dielectric}  & 9.3\cite{weeks1922dielectric}   & 6.4\cite{weeks1922dielectric}   
& $10^{-14}$\cite{micapargroup}           & $5 \times 10^{-16}$$^{\dagger}$   & $2.5 \times 10^{-13}$$^{\dagger}$ \\

PAN          & 0.021     & 3.54$^{\ddagger}$ & 4.2\cite{haynes2016crc}   & 2.87\cite{haynes2016crc}  
& $10^{-13}$$^{\ddagger}$ & $10^{-15}$\cite{parinov2015advanced}          & $10^{-11}$\cite{haynes2016crc} \\

SiO$_2$      & 3$^{\dagger}$         & 3.8$^{\ddagger}$   & 3.9\cite{el2012fundamentals}   & 3.7\cite{el2012fundamentals}    
& $10^{-15}$\cite{el2012fundamentals}     & $10^{-16}$\cite{shackelford2000crc}           & $10^{-13}$ \cite{palumbo2020review} \\
\hline\hline
\end{tabular}
\end{table*}

\section{Charge decay with a variety of materials}
Here we present further data for a variety of materials with different nominal conductivities. Table \ref{tab:material_properties} shows the relevant parameters for all samples used. Note that, as discussed in the main text, the non-Ohmic nature of the materials leads to nominal values for conductivity that can vary over several orders of magnitude for a given material (last two columns in table).  

\begin{figure}[!ht]
\centering
\includegraphics{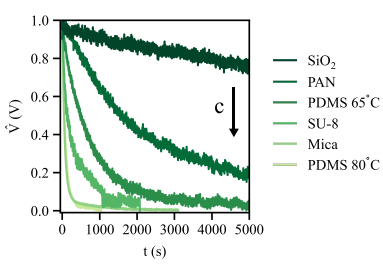}
\caption{ \label{fig:SI_fig_decays} Better insulators decay more slowly. When we unfold spatial KPFM data into the time domain for many materials of varying conductivities, we see that the `best' insulators (\textit{e.g.}~SiO$_2$) decay slowly, while the `worst' ones (\textit{e.g.}~PDMS) decay quickly. Potentials are normalized to initial values aid visualization.}
\end{figure}

Suppl.~Fig.~\ref{fig:SI_fig_decays} shows measured KPFM potentials for each material after contact, unfolded from the space domain to the time domain. All data is normalized to the initial measured potential for visual aid. For the `best' insulator (SiO$_2$), the decay appears linear over the timescale shown; over longer timescales its decay slows down. As the nominal conductivities increase, the decays clearly occur more quickly, with the fastest corresponding to the `worst'  insulator, PDMS.

\section{Cole-Cole response function and fit parameters}
\begin{figure}[!ht]
\centering
\includegraphics{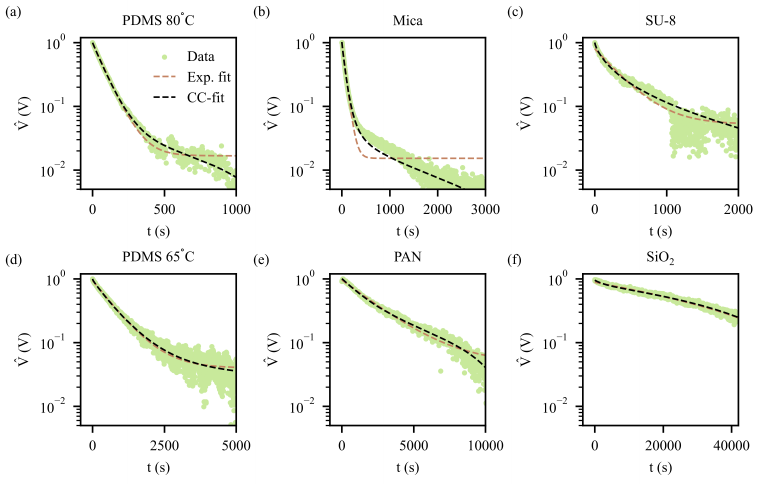}
\caption{ \label{fig:SI_fig_cc_fit} Comparison between exponential fit and Mittag-Leffler fit. (a)-(f) We compare the exponential model (Exp. fit) and the response function model (CC-fit) for all used materials on log-lin plots. Normalized potentials are shown for visual aid. In all cases, Cole-Cole response model fits as well or better.  Fit values are listed in Suppl.~Tables \ref{tab:exponential parameters} and \ref{tab:cole-cole parameters}, respectively.}
\end{figure}

As we explained in the main text, the measured potential decays are not strictly exponential; this is highlighted by plotting them on log-lin scale, as in Suppl.~Fig.~\ref{fig:SI_fig_cc_fit}.  Ref.~\cite{molinie2023dielectric} developed a more rigorous approach to model such decay considering as a circuit composed of several series cells, each corresponding to an elementary relaxation process. In this framework, a "cell" corresponds to a discrete relaxation mode within the material, capturing contributions from different polarization or charge transport mechanisms. The resulting expression found in that paper is,
\begin{equation}
    V(t) = V_0 - \sum_{i=1}^2 \Delta V_i [1 - E_{\alpha_i}(-(t/\tau_i)^{\alpha_i})]
    \label{eq:cole-cole decay}
\end{equation}
with $\alpha_i$, $\tau_i$, $\Delta V_i$ and $E_{\alpha_i}$ being characteristic exponent, relaxation time, voltage change and Mittag-Leffler function associated with the $i^{th}$ cell, respectively. Following Ref.~\cite{molinie2023dielectric} two cells are sufficient to fit the relaxation dynamics in our data. We use Eq.~\ref{eq:cole-cole decay} to fit the experimental data (green dots) and find good agreement, as shown by the black dashed line in Suppl.~Fig.~\ref{fig:SI_fig_cc_fit}.

\begin{table}[ht!]
\centering
\caption{\parbox[t]{\textwidth}{\raggedright Fit parameters used for the exponential function: initial potential $V_{0}$, characteristic time constant $\tau$ and the \newline
back ground potential $V_{bg}$ \newline \vspace{1mm}}}
\label{tab:exponential parameters}
\begin{tabular}{lccc}
\hline\hline
Material & $V_0$ (V) & $\tau$ (s) & $V_{bg}$ (V)\\
\hline
SU-8          & 0.75   & 337  & 0.05   \\
PDMS 65$^{\circ}$C & 0.9    & 740   & 0.04   \\
PDMS 80$^{\circ}$C & 0.96   & 86   & 0.02   \\
Mica          & 1.0    & 70   & 0.02   \\
PAN           & 0.92   & 2482 & 0.05   \\
SiO$_2$          & 1.35   & 65105 & -0.46 \\
\hline\hline
\end{tabular}
\end{table}

\begin{table}[ht!]
\centering
\caption{\parbox[t]{\textwidth}{\raggedright Fit parameters used for the Cole-Cole response function: initial potential $V_{0}$, potential decay contribution of \newline
corresponding cell $\Delta V$, characteristic time constant $\tau$ and characteristic exponent $\alpha$    \newline \vspace{1mm}}}
\label{tab:cole-cole parameters}
\begin{tabular}{lccccccc}
\hline\hline
Material & $V_0$ (V) & $\Delta V_1$ (V) & $\Delta V_2$ (V) & $\tau_1$ (s) & $\tau_2$ (s) & $\alpha_1$ & $\alpha_2$ \\
\hline
SU-8          & 1.0    & 1.05   & 0    & 255  & 4591  & 0.73   & 0.31   \\
PDMS 65$^{\circ}$C & 1.0    & 0.1    & 0.9  & 441  & 720   & 0.36   & 0.97   \\
PDMS 80$^{\circ}$C & 1.0    & 0.97   & 0.15 & 81   & 5672  & 0.98   & 1.0    \\
Mica          & 1.02   & 0.99   & 0.07 & 64   & 4736  & 0.94   & 0.45   \\
PAN           & 1.01   & 0.73   & 1.59 & 1612 & 62035 & 1.0    & 1.0    \\
SiO$_2$          & 0.95   & 0.13   & 2.54 & 2639 & 163213 & 0.96   & 1.0    \\
\hline\hline
\end{tabular}
\end{table}

\newpage

\bibliographystyle{apsrev4-2}
\bibliography{manuscript}